\documentclass[%
 aip,
 amsmath,amssymb,
 reprint,%
aps
]{revtex4-1}

\usepackage{cancel}
\usepackage{braket}
\usepackage{hyperref}
\usepackage{adjustbox}
\usepackage{float}
\usepackage{amsmath}
\usepackage{multirow}

\usepackage{graphicx}
\usepackage{dcolumn}
\usepackage{bm}

\usepackage[utf8]{inputenc}
\usepackage[T1]{fontenc}
\usepackage{mathptmx}
\usepackage{etoolbox}
\usepackage{xcolor}
\usepackage{color}
\usepackage{soul}

\hypersetup{
    colorlinks=true,
    linkcolor=blue,
    }

\newcommand\eq[1]{
\begin{equation*}
    #1
\end{equation*}}

\newcommand\numeq[1]{
\begin{equation}
    #1
\end{equation}}

\newcommand\Airy{\mathsf{Ai}}
\newcommand\Bairy{\mathsf{Bi}}
\newcommand\lin{\text{Lin}}

\begin{document}

\preprint{AIP/123-QED}

\title{The Feynman path integral formulation of non-dispersive Airy wave packets and their applications to the heavy meson mass spectra and ultra-cold neutrons}

\author{Paul Ferrante}

\author{Connor Donovan}

\author{Chueng-Ryong Ji}
\noaffiliation

\affiliation{Department of Physics, North Carolina State University, Raleigh, North Carolina 27695-8202, USA}

\date{\today}

\begin{abstract}
We demonstrate the non-spreading behavior of Airy wave packets utilizing the Feynman path integral formulation of a linear potential, the Airy functions' zeros correspondence to heavy-meson mass spectroscopy, and their implications to the eigenstates of ultra-cold neutrons in Earth's gravitational field. We derive the linear kernel, and utilize the Feynman path integral time evolution to show that Airy function wave packets are non-dispersive in free space. We then model the confining contribution to 1S-2S heavy meson mass gaps as a 1+1D absolute linear potential and look at the correspondence of the Airy function zeros. In doing so, we predicted the confining contribution to the mass gap of heavy mesons with a good accuracy when compared to calculations performed in the light front. Furthermore, we used these Airy function solutions to model the quantum states of a neutron under Earth’s gravity. We show that the measured heights of a neutron can be modeled by the zeros of the Airy function, and compare to experimental data and predictions utilizing the WKB approximation.
\end{abstract}

\maketitle

\section{\label{Intro}Introduction \protect}

The Airy function, which is the solution to the Stokes equation, has been found to have a myriad of physical and mathematical applications, ranging from optics and rainbows to fluid dynamics and probability. The hyperasymptotic nature of the Airy function even gives it applications to the Chern-Simons theory\citep{Witten}. It is relevant when describing a number of quantum systems under a constant force, which corresponds to linear potentials. In addition to these phenomena, the Airy function eigenstates of linear potentials have the intriguing property that they are non-dispersive when left to propagate in free space.A method by which one can prove this is the Feynman path integral, which propagates wave functions through time by integrating over all possible paths a particle can take when moving between points. 

We explore the Feynman path integral formulation of the Airy wave function eigenstates of a linear potential to demonstrate their non-spreading behavior when let free, as discussed by M. V. Berry and N. L. Balazs \citep{wavepackets}. We will investigate these eigenstate solutions and demonstrate their utility in modeling certain physical systems. As the Airy functions are the exact solution to the time-independent Schr{\"o}dinger equation with a linear potential, they can be utilized for understanding heavy-meson mass spectroscopy and ultra-cold neutrons bouncing in Earth's gravitational field. Corresponding the Airy function zeros to a simple 1+1D model of heavy-meson mass spectroscopy gives insight into the confining potential's contribution to their mass gaps. Application of the Airy function solutions to modeling to the quantum behavior of ultra-cold neutrons in a gravitational field demonstrates a clear difference from the prediction of the classical mechanics as the measured heights are quantized by the Airy function zeros.

In Section \ref{Linear}, we derive the Airy wave function in scaled coordinates. In Section \ref{Feynman Path Integral}, we go over the general theory and derivations of the Feynman path integral formulation, from which we will show that eigenstates of linear potentials are non-dispersive in free potentials. Next in Section \ref{ABS and One Side}, we derive the Airy eigenstates of different variations of linear potentials. Using these alternative linear potentials, we then discuss the applications of these formulations to 1S-2S mass gaps in the heavy-meson sector in Section \ref{Quark}, and to the bouncing of ultra-cold neutrons on Earth's surface in Section \ref{Neutron}. Following this is our conclusion and outlook in Section \ref{Conclusion}. We end with the appendices, where we show a derivation of the Airy function in Appendix \ref{appen b}, outline the theory of the kernel derivation in Appendix \ref{appen e}, show the derivation of the linear propagator in momentum space in Appendix \ref{appen c}, and finally we show how the differences in the zeros of the Airy function approaches 0 in the WKB approximation in Appendix \ref{appen d}. 

\section{\label{Linear}1-D Linear Potential Solutions \protect}

A linear potential in 1 dimension has the form $V(x) = 
Fx$. Here, $F$ is the magnitude of the force, analogous to the tension of a string. The eigenstates that describe this potential are then the solutions to 
\numeq{-\frac{\hbar^2}{2m}\frac{\partial^2}{\partial x^2}\psi + Fx \psi = E\psi. }
Scaling the Schr\"{o}dinger equation by the characteristic length of the system,

\numeq{ x_{0}= \left( \frac{\hbar^2}{2mF} \right)^{1/3}, \label{rescale}}
 results in the dimensionless Schr\"{o}dinger equation 

\numeq{\frac{\partial^2}{\partial q^2} \psi = (q - E_q)\psi, \label{Stokes}}
where the dimensionless scaled position coordinate $q$ and energy eigenstates $E_{q}$ are defined as

\numeq{q=\frac{x}{x_{0}} \text{ , } E_q = \frac{E}{Fx_{0}}.}
Note here that the position and energy appear on par in terms of their corresponding dimensionless variables $q$ and $E_q$, respectively, in Eq. (\ref{Stokes}). 
As we will see, this equal footing feature of $q$ and $E_q$ provides the spacing of energy eigenvalues distinctively different from nonlinear potential systems. As Eq. (\ref{Stokes}) takes the form of a Stokes equation, its solution can be given by

\begin{figure}[h]
    \centering
    \includegraphics[width=8.5cm,height=8.5cm,keepaspectratio]{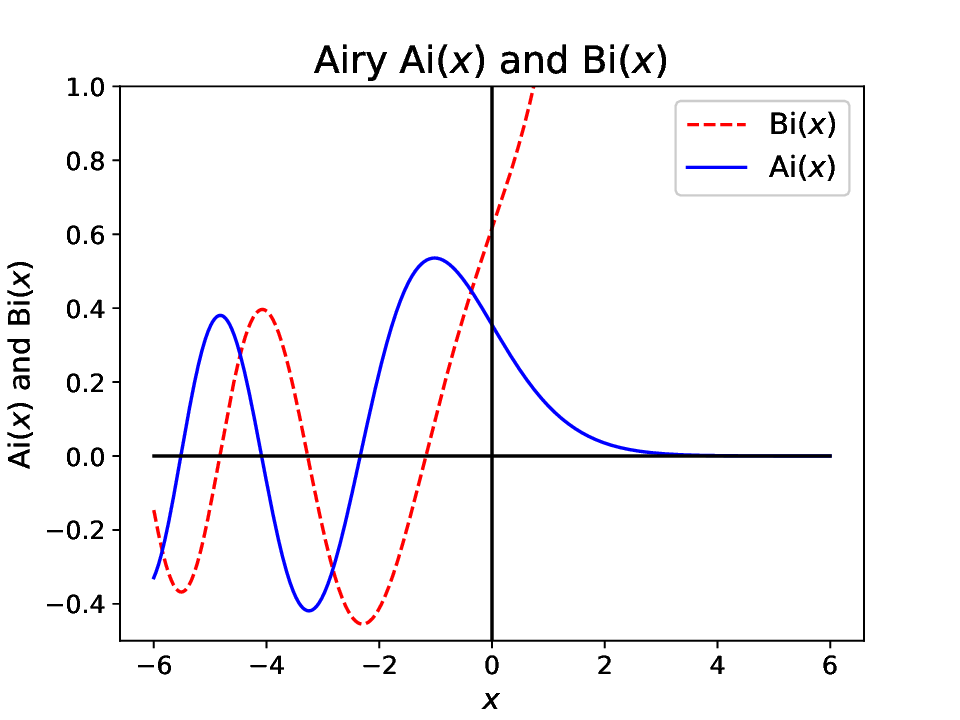}
    \caption{Graph of Airy $\Airy(x)$ and $\Bairy(x)$ functions. }
    \label{Ai and Bi}
\end{figure}

\numeq{\psi(q) = C \Airy(q- E_q) + D\Bairy(q- E_q),}
where $\Airy$ and $\Bairy$ represent the linearly independent Airy functions, as shown in Fig. \ref{Ai and Bi}. Noting that the wave function must terminate as $q$ approaches infinity, $D$ must be 0 as the $\Bairy$ function diverges. This results in

\numeq{\psi(q) = C \Airy(q- E_q). \label{lin func}}

\section{\label{Feynman Path Integral} Feynman Path Integral \protect}

\subsection{\label{Kernel}Linear Kernel \protect}

In order to evolve a wave function from some initial time and position $(x', t')$ to some later time and position $(x, t)$ by means of the Feynman path integral formulation, one must first find the kernel of the system $K(x,t;,x',t')$. Once this is known, the time evolved state can be calculated as 
\numeq{ \psi(x,t)= \int^{\infty}_{-\infty} dx'K(x,t;x',0)\psi(x',0),}
where $\psi(x',0)$ is the initial wave function at $t'=0$. One can calculate the kernel of a system using the time-slice method, which involves summing over infinitesimally small time intervals or "slices" of time. As shown in Appendix \ref{appen e}, we carry out this calculation starting from the following expression for the kernel:
\begin{align}
    K(x,t;x',t') & = \lim_{N \to \infty}\int dx_{N-1} \dots dx_{1}(\frac{m}{2 \pi i \hbar \epsilon })^{N/2} e^{\frac{i}{\hbar}S[x(t)] } \nonumber \\
    & = \int_{x(t')}^{x(t)} \mathcal{D}[x(t)] e^{\frac{i}{\hbar}S[x(t)] },
\end{align}
where we define 

\eq{\mathcal{D}[x(t)]=\lim_{N \to \infty} \left(\frac{m}{2 \pi i \hbar \epsilon }\right)^{N/2}dx_{N-1} \dots dx_{1},} 
with $ \epsilon=(t-t')/N$. Since $\hbar$ is small, there will be rapidly oscillating phases that render most trajectories unstable and cause them to cancel. The exception is the classical trajectory, which survives satisfying the least action principle in the path integral. This allows us to make use of the stationary phase approximation such that only the stationary phase, or classical path, and it's quantum fluctuations survive, leaving us with Eq. (\ref{appen J}):

\numeq{ K(x,t;x',t') = J(t,t')e^{\frac{i}{\hbar} S[x_{\text{cl}}(t)]}, \label{J(t)}}
where $J(t, t') = K(0, t; 0, t')$ is a purely time dependent prefactor as $\delta x'= \delta x =0$ on the endpoints. \\
The classical action for a linear potential is given by

\begin{align}
   S_{\lin}(x,t;x',t') = \frac{m}{2} & \left( \frac{(x - x')^2}{t-t'} - \frac{F }{m}(t-t')(x + x') \right. \nonumber \\
    & \left. - \frac{F^2}{12m^2}(t-t')^3 \right). \label{action}
\end{align}
Thus, the kernel for the linear potential can be written as 

\begin{align}
    K_{\lin}(x,t;x',t')=J(t,t')e^{\frac{i}{\hbar} S_{\lin}(x,t:x',t')} \label{pracprop}
\end{align}
Our next task is to find the functional form for $J(t,t')$. To do this, we will implement a strategy used by B. Holstein where we introduce an intermediate position and time $(x'',t'')$ to split our time evolution into two steps\citep{Holstein}. Doing so results in 

\eq{\bra{x}e^{\frac{-i}{\hbar}Ht}\ket{x'} = \bra{x}e^{\frac{-i}{\hbar}H(t-t'')} e^{\frac{-i}{\hbar}H(t''-t')} \ket{x'}, }
and by inserting the identity, we obtain

\numeq{\bra{x}e^{\frac{-i}{\hbar}Ht}\ket{x'} = \int_{-\infty}^\infty dx'' \bra{x}e^{\frac{-i}{\hbar}H(t-t'')} \ket{x''}\bra{x''} e^{\frac{-i}{\hbar}H(t''-t')} \ket{x'}. \label{completeness}}
Now taking $\Delta t = t-t''$ and substituting Eq. (\ref{pracprop}) into Eq. (\ref{completeness}) gives

\begin{align}
    J(t,0)e^{\frac{i}{\hbar} S_{\lin}(x, t ; 0, 0)} & = J(t,t'')J(t'',0) \nonumber \\
    &  \times \int_{-\infty}^{\infty}dx'' e^{\frac{i}{\hbar}S_{\lin}(x, t, x'', t'')} e^{\frac{i}{\hbar} S_{\lin}(x'', t'' ; 0, 0)},
\end{align}
where we took our initial position and time as $(x',t')=(0,0)$ for simplicity. Substituting $S_{\lin}$ from Eq. (\ref{action}) and factoring terms not dependent on $x''$ from our integral and moving them to the left side of the equation yields

\begin{widetext}
\begin{align}
    \frac{J(t,0)}{J(t'',0)J(t,t'')} \exp \left[ \frac{im }{2\hbar} \left( x^2 \left( \frac{1}{t} - \frac{1}{\Delta t} \right) + \frac{F^2 }{4m^2} ( t^2t''-tt^2) - \frac{Fx t''}{m} \right) \right] & = \int^{\infty}_{-\infty} dx'' \exp\left[\frac{im}{2\hbar} \left( x''^2 \left( \frac{1}{\Delta t} + \frac{1}{t''} \right) - x''\left( \frac{2x}{\Delta t} + \frac{Ft}{m} \right) \right)\right]. \label{precube}
\end{align}
\end{widetext}
This integral can be evaluated by taking advantage of the integral formula given in Eq. (\ref{useful integral}), and in doing so we find

\numeq{\frac{J(t,0)}{J(t'',0)J(t,t'')} = \sqrt{\frac{2\pi i \hbar \Delta tt''}{mt}}. }
Solving for the functional form of $J(t,t)$ then leads to

\numeq{J(t,t') = \sqrt{\frac{m}{2\pi i \hbar (t-t')}}.}
We arrive at the final expression for the linear kernel as

\begin{align}
    K_{\lin}(x, t ; x', t') & = \sqrt{\frac{m}{2\pi i \hbar (t-t')}} \exp\left[ \frac{im}{2\hbar} \left(  \frac{(x - x')^2}{t-t'} \right. \right. \nonumber \\ 
    & \left. \left.- \frac{F }{m}(t-t')(x + x')  - \frac{F^2}{12m^2}(t-t')^3 \right) \right], \label{klin}
\end{align}
in agreement with H. Kleinhart's solution\citep{Kleinhart}. 

An important observation to note is that the prefactor in this case is equal to the prefactor in the free case. In taking the strength of the linear potential $F$ to be zero, we see that 

\begin{align}
    K_{\lin}(x, t ; x', t')|_{F=0}
    & = J(t, t')\exp\left[ \frac{im}{2\hbar} \frac{(x - x')^2}{t-t'}  \right] \nonumber \\
    & = J(t, t')\exp\left[ \frac{i}{\hbar}S_{\text{Free}} \right] \nonumber \\
    & = K_{\text{Free}}(x, t ; x', t'). \label{kfree}
\end{align}
This is because the linear kernel is, in fact, a generalization of the free kernel. Rewriting ${K_{\lin}}$ in Eq. (\ref{klin}) by isolating the second and third terms in the exponentiated $S_{\text{Lin}}$ and substituting in $K_{\text{Free}}$ as it's shown in Eq. (\ref{kfree}) results in

\begin{align}
    & K_{\lin}(x, t ; x', t') =  K_{\text{Free}}(x, t ; x', t') \nonumber\\
    & \times \exp\left[ \frac{im}{2\hbar} \left(- \frac{F}{m}(t-t')(x + x')  - \frac{F^2}{12m^2} (t-t')^3\right) \right]. \label{Kx}
\end{align}
This form of the linear kernel clearly shows how it is a generalization of the free kernel.

\subsection{\label{propagation} Non-Dispersal in Free Potential}

A unique property of these Airy wave packets is that when placed in a free potential, $V(x) = 0$, they do not dissipate over time\citep{wavepackets}. The Airy wave packets instead undergo position translation with the probability density shape remaining invariant with time. This can be intuitively understood to occur because the Airy function is a generalization of the freely propagating wave, but here we provide a quantitative proof of this behavior utilizing Feynman path integral time evolution. We find it useful in the subsequent calculation to implement again the coordinate rescaling defined in Eq. (\ref{rescale}) and rewrite the free kernel as

\numeq{K_{Free(q,t:q',t')}=\left( \sqrt{\frac{m x_{o}^2}{2 \pi i \hbar (t-t')}} \right) \exp\left[\frac{i m x_{o}^{2}(q-q')^{2}}{2 \hbar (t-t')}\right].\label{freeprop}} 
The momentum representation of the quantum mechanical Airy function is given by substituting the dimensionless variables $k=px_0 / \hbar$ and $q=x/x_0$ in Appendix \ref{appen b}. Taking our wave function at the initial time $t'=0$ to be the Fourier transformation of the Airy function shifted by the scaled energy eigenstates $E_q$, it becomes 

\begin{align}
    \psi(q, 0) & = \Airy(q-E_q) \nonumber \\
    & = \frac{1}{\sqrt{2\pi \hbar}}\int^{\infty}_{-\infty} dp \exp\left[\frac{i p^{3}x_{o}^{3}}{3 \hbar^3 }\right]\exp\left[\frac{ip x_{o}}{\hbar } (q-E_q)\right],
\end{align}
for which it is important to notice that the initial momentum space wave function takes the form

\numeq{\phi(p,0) = \exp\left[\frac{i p^{3}x_{o}^{3}}{3 \hbar^3 }\right]\exp\left[\frac{ip x_{o}}{\hbar } (q-E_q)\right]. \label{mswf}}
Inputting this into the Feynman path integral time evolution of a free potential results in

\begin{align}
    \psi(q,t)& = \frac{1}{\sqrt{2\pi \hbar}} \sqrt{\frac{m x_{o}^2}{i \hbar t}}\int^{\infty}_{-\infty}dq'\int^{\infty}_{-\infty} dp \text{ }\exp\left[\frac{i m x_{o}^{2}(q-q')^{2}}{2 \hbar t}\right] \nonumber \\
    & \times \exp\left[\frac{i p^{3} x_{o}^{3}}{3 \hbar^3 }\right] \exp\left[\frac{ip x_{o}}{\hbar } (q'-E_q)\right].
\end{align}
 Expanding the $(q-q')^2$ term and carrying out the $q'$ integration yields

\begin{align}
    \psi(q,t)&=\frac{1}{\sqrt{2\pi \hbar}} \int^{\infty}_{-\infty} dp \exp\left[i\left(\frac{p^{3} x_{o}^{3}}{3 \hbar^{3} } \right. \right. \nonumber \\
    & \left. \left. -\frac{ p^{2} t}{2 m  \hbar}\right)\right] \exp\left[\frac{ip  x_{o}}{\hbar} (q-E_q)\right]. \label{precube}
    \end{align}
Now we can complete the cube in the first exponential\citep{Golub}, which results in
\begin{align}
    &\psi(q,t) =\frac{1}{\sqrt{2\pi \hbar}} \int^{\infty}_{-\infty} dp \exp\left[i\left(\frac{1}{3}\left( \frac{p x_{o}}{\hbar}-\frac{\hbar t}{2m x_{o}^{2}}\right)^{3} \right. \right. \nonumber \\
    & \left. \left. -\left( \frac{p x_{o}}{\hbar}\right)\left(\frac{\hbar t}{2 m x_{o}^{2}}\right)^{2}+
    \frac{1}{3}\left(\frac{\hbar t}{2 m x_{o}^{2}}\right)^{3}\right)\right] \exp\left[\frac{i p x_{o}}{\hbar} (q-E_q)\right].
\end{align}
Denoting the factor 

\eq{\frac{u x_0}{\hbar} = \left ( \frac{p x_0}{\hbar}-\frac{\hbar t}{2mx_0^2}\right ),}
we get that

\begin{align}
    &\psi(q,t) = \frac{1}{\sqrt{2\pi \hbar}} \exp\left[ \frac{i \hbar t}{2mx_{o}^2}\left( \left (q-E_q \right ) - \frac{2}{3}\left( \frac{\hbar t}{2mx_{o}^2}\right)^2  \right) \right] \nonumber \\
    & \times \int_{-\infty}^\infty du \exp\left[ \frac{iu^3 x_{o}^3}{3 \hbar^3} \right] \exp\left[  \frac{iu x_{o}}{\hbar}\left( (q-E_q) - \left(\frac{\hbar t}{2mx_{o}^2}\right)^2 \right) \right]. 
\end{align}
The term inside the integral is then nothing but the Fourier transform definition of our original Airy function shifted by a quadratic time term, modulo a non-physical phase factor:

\begin{align}
    \psi(q,t) & = \Airy \left( (q-E_q)-\left( \frac{\hbar t}{2 m x_{o}^{2}}\right)^{2}  \right) \nonumber \\ 
    & \times \exp\left[ \frac{i \hbar t}{2mx_{o}^2}\left( \left (q-E_q \right ) - \frac{2}{3}\left( \frac{\hbar t}{2mx_{o}^2}\right)^2  \right) \right], \label{tevolved}
\end{align}
in excellent agreement with M. V. Berry and N. L. Balazs\citep{wavepackets}. \\

One can perform an equivalent calculation utilizing the momentum space representation of Eq. (\ref{freeprop}) as derived in Appendix \ref{appen c};

\numeq{K_{Free(p,t:p',t')}=\delta\left( p-p'\right) \exp\left[\frac{i p'^{2} (t-t') }{2 \hbar m }\right].}
In using this to time evolve our momentum space wave function Eq. {\ref{mswf}}, we find

\begin{align}
    \phi(p,t)& =\int^{\infty}_{-\infty}dp' K_{Free(p,t:p',0)} \phi(p,0) \nonumber \\
    & = \exp\left[\frac{i p^{3} x_{o}^{3}}{3 \hbar^3 } - \frac{i p^{2} t }{2 \hbar m }\right] \exp\left[\frac{ip x_{o}}{\hbar } (q-E_q)\right],
\end{align}
consistent with Eq.(\ref{precube}). \\

To show a zero dispersion, we can calculate the variance in $q$. Using Eq. (\ref{tevolved}), we can show that the time-dependent expectation value of $q$ is 

\begin{align}
\braket{q}_{t}& =\int_{-\infty}^{\infty} q \psi(q,t)^{*}\psi(q,t)dq \nonumber \\ 
& = \int_{-\infty}^{\infty} q \Airy\left((q-E_q)-\left( \frac{\hbar t}{2 m x_{o}^{2}}\right)^{2}\right)^{2}dq. \label{<q>}
\end{align}
We make a substitution such that 

\eq{q'= q-\left( \frac{\hbar t}{2 m x_{o}^{2}}\right)^{2},}
and in substituting this into Eq. (\ref{<q>}), we find

\begin{align}
\braket{q}_{t}& =\int_{-\infty}^{\infty} \left( q'+\left( \frac{\hbar t}{2 m x_{o}^{2}}\right)^{2} \right) \Airy(q' - E_q)^2dq' \nonumber \\ 
& = \braket{q}+\left( \frac{\hbar t}{2 m x_{o}^{2}}\right)^{2}. 
\end{align}
Following a similar method for the time-dependent expectation value of $q^2$, we find
\begin{align}
\braket{q^2}_{t}&= \int_{-\infty}^{\infty} \left( q'+\left( \frac{\hbar t}{2 m x_{o}^{2}}\right)^{2} \right)^{2} \Airy(q' - E_q)^2 dq'  \nonumber \\
&=  \braket{q^2} + 2 \left( \frac{\hbar t}{2 m x_{o}^{2}}\right)^{2} \braket{q} + \left( \frac{\hbar t}{2 m x_{o}^{2}}\right)^{4}.
\end{align}
The dispersion of a function is represented by $\Delta q^2_{t} =\braket{q^2}_{t}-\braket{q}^{2}_{t}$, and in performing the difference we find that all time dependent terms will in fact cancel, which implies that the variance of the Airy wave packet in $q$ does not disperse with time. Taking the modulus of Eq. (\ref{tevolved}), we get the following probability density.

\begin{figure*}
\includegraphics[width=17cm, keepaspectratio]{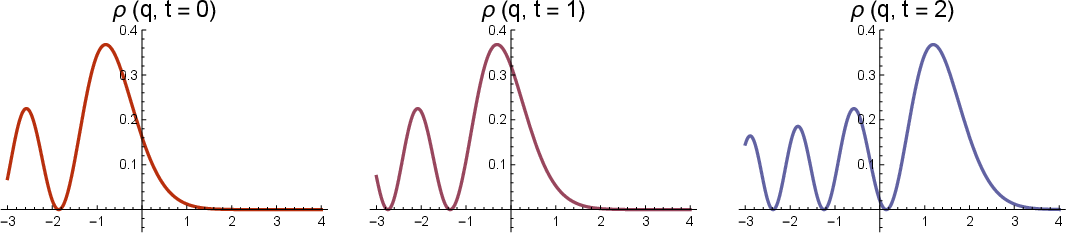}
\caption{The probability density $\rho(q=x/x_0,t)$ of an Airy wave packet at 3 different times. The probability density shows no dispersion, but instead accelerates along the $q$ axis. Here, $t$ is in units of $2m x_0^2/\hbar$.  }
\label{Airy Graphs}
\end{figure*}

\numeq{|\psi(q,t)|^{2}=\Airy^2 \left( (q-E_q)-\left( \frac{\hbar t}{2 m x_{o}^{2}}\right)^{2} \right).}
This shows that when the Airy wave packet is released in a free potential, its probability density $\rho(q,t) = |\psi(q,t)|^2$ will not disperse, but instead propagates (actually accelerates) through space as time increases, as depicted in Fig \ref{Airy Graphs}.

\section{\label{ABS and One Side} Eigenstates of alternative linear potentials \protect}

\subsection{\label{ABS} Absolute Linear Potential }

An absolute linear potential is given by $V(x) = F|x|$. Inputting this into the time independent Schr{\"o}dinger equation with the same coordinate rescale as earlier in Eq. (\ref{rescale}) and solving leads us to  

\numeq{\psi(q) = C \Airy(|q|- E_q) + D \Bairy(|q|- E_q). }
Since quarks are not found outside of confinement in the low energy regime, the wave function must terminate as $|q|$ approaches infinity. This once again corresponds to $D$ equaling 0 as the $\Bairy$ function diverges. Similarly to Eq. (\ref{lin func}), this results in

\begin{figure}[h]
    \centering  \includegraphics[width=8.5cm,height=8.5cm,keepaspectratio]{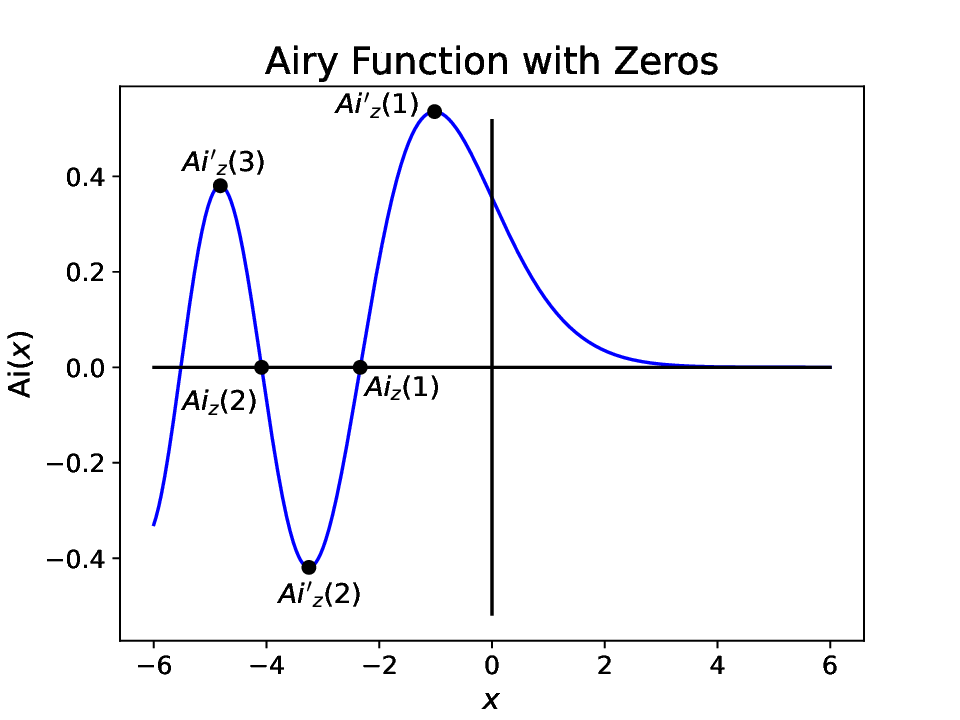}
    \caption{Graph of $\Airy(x)$, with marked zeroes for $\Airy(x)$ and $\Airy'(x)$.}
    \label{Zeros}
\end{figure}

\numeq{\psi(q) = C \Airy(|q|- E_q).}
The energy eigenstates can be found for these solutions by examining the other boundary conditions. For even functions, the wave functions must satisfy $\psi'(0) = 0$, and so 
\numeq{\psi'(0) = 0 = C\Airy'(-E_q). \label{absWave}}
Here, $\Airy'$ denotes the first derivative of the $\Airy$ function. We can see that Eq. (\ref{absWave}) is satisfied when $-E_q$ is a zero of the $\Airy'$ function. The zeros of $\Airy$ and $\Airy'$ are well known, and we will denote their $n^{th}$ zero as $\Airy_z(n)$ and $\Airy'_z(n)$, respectively, as shown in Fig. \ref{Zeros}. The energy eigenvalues for the even states are then given by 
\numeq{E_{qn} = -\Airy'_z(\frac{n}{2}+1), }
where $n$ must be even.
Odd functions must satisfy $\psi(0) = 0$, which means that
\eq{\psi(0) =0= C\Airy \left(-E_q \right).}
Just like before, this is satisfied when the energy value is a zero of $\Airy$. Solving for the odd energy values results in 
\numeq{E_{qn} = -\Airy_z(\frac{n-1}{2} + 1),}
where n is odd here.
This means that the allowed energy eigenvalues of this system are 

\numeq{E_{qn} = \begin{cases}
    -\Airy'_z(\frac{n}{2}+1) \text{ , } n \text{ is even}, \\
    -\Airy_z(\frac{n-1}{2} + 1) \text{ , } n \text{ is odd} .\\
\end{cases} \label{q Energies}}
Since the zeros of the $\Airy$ and $\Airy'$ functions are negative, all the allowed energy values will be positive. Undoing the coordinate transformation gives the following result.
\numeq{E_n = \begin{cases}
    -\sqrt[3]{\frac{\hbar^2 F^2}{2 m}}\Airy'_z(\frac{n}{2}+1) \text{ , } n \text{ is even}, \\
    -\sqrt[3]{\frac{\hbar^2 F^2}{2 m}}\Airy_z(\frac{n-1}{2} + 1) \text{ , } n \text{ is odd}. \\
\end{cases} \label{Energies}}
We can see that as the value of $F$ increases, the energy eigenvalues should increase as well, but as the mass increases, the energy eigenvalues will decrease. We can normalize the wave function to find the following.
\eq{C_n = \sqrt{\frac{1}{2[ E_{qn}\Airy ^2(- E_{qn} ) + \Airy'^2(- E_{qn} )]}}, }
where $C_n$ denotes the $n^{th}$ normalization constant. Substituting the energies from Eq. (\ref{q Energies}), this becomes
\numeq{C_n = \begin{cases}
    \sqrt{\frac{1}{2E_{qn}\Airy ^2(-\Airy'_z(\frac{n}{2} + 1)) }} \text{ , } n \text{ is even}, \\
    \sqrt{\frac{1}{2 \Airy'^2(-\Airy_z(\frac{n-1}{2}+1))}} \text{ , } n\text{ is odd}. \label{abs norm}
\end{cases}}
We can see that switching between even and odd parities automatically removes a component from the normalization constant. This means that the eigenfunctions can be written universally for both even and odd parity as 
\numeq{\psi_n(x) = C_n\Airy (|q| - E_{qn}).}
%
 \begin{figure}[h]
    \centering
    \includegraphics[width=8.5cm,height=8.5cm,keepaspectratio]{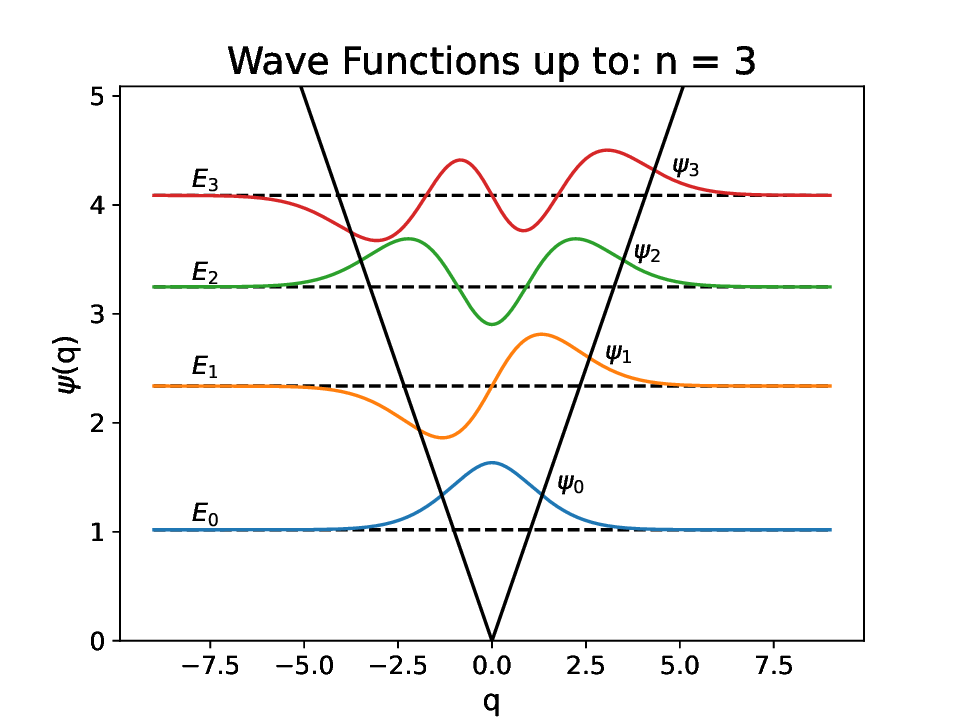}
    \caption{The first four eigenstates of the absolute linear potential, scaled to prevent overlap.}
    \label{fig:1.7}
\end{figure}

\subsection{\label{One Side} One Sided Potential Potential }

A one-sided linear potential has the form
\numeq{V(x) = \begin{cases}
    \infty & x < 0,\\
    Fx & x \geq 0. \\
\end{cases}}
The process of solving this system is exactly as in the absolute value case; however, the boundary conditions change. $\psi(x)$ must terminate at the left boundary, so $\psi(x)=0$ for all $x\leq 0$. This boundary condition terminates all even-parity solutions, and we are only left with solutions of odd parity. Thus, for $q\geq 0$, 
\numeq{\psi_n(x) =C_n\Airy (q - E_{qn}) \text{ , } }
where
\numeq{E_{qn} = -\Airy_z(n + 1) \label{one side energies}}
and
\numeq{C_n =\sqrt{\frac{1}{ \Airy'^2(-E_{qn}) }}.}
This is the same as Eq. (\ref{abs norm}) with the normalization constant for the odd absolute potential solution, but without the factor of $\sqrt{1/2}$, since the one-sided wave function will occupy half the space of the absolute value wave function.

 \begin{figure}[h]
    \centering
    \includegraphics[width=8.6cm,height=8.5cm,keepaspectratio]{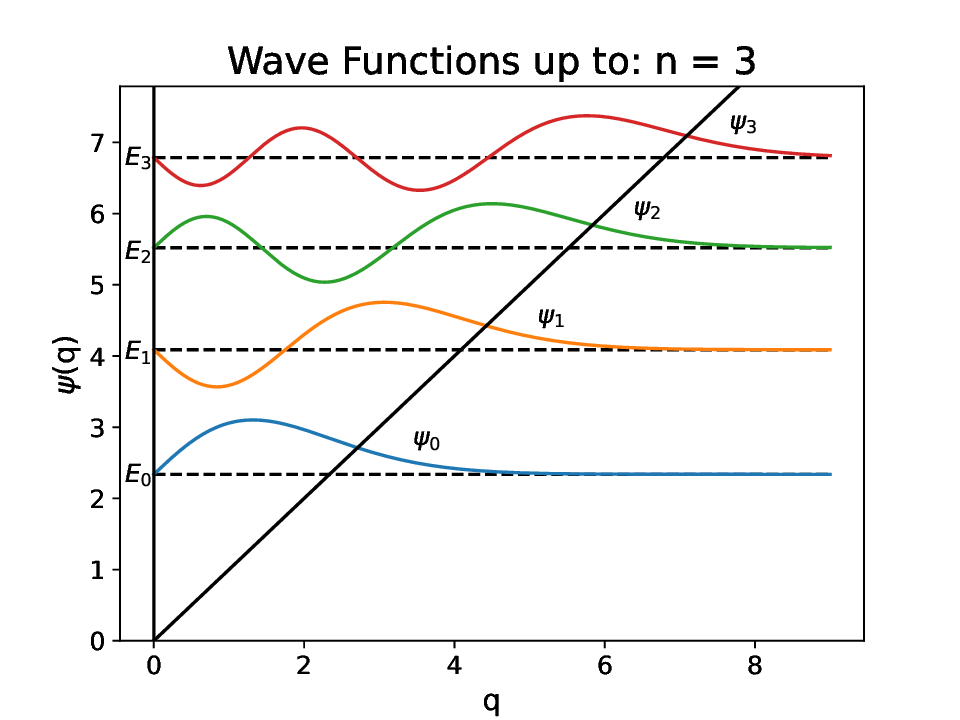}
    \caption{The first four eigenstates of the one sided linear potential, scaled to prevent overlap.}
    \label{fig:1.7}
\end{figure}

\section{\label{Quark} Application to the heavy meson mass spectra \protect}

Data from the particle data group show that the 1S-2S mass gaps of heavy mesons are flavor-independently 600 MeV \citep{Jafar, PDG}. This is an interesting phenomenon, and we want to utilize the Airy functions to explain the confining contribution of these 600 MeV mass gaps.

\begin{figure}[h]
    \centering   \includegraphics[width=8.5cm,height=8.5cm,keepaspectratio]{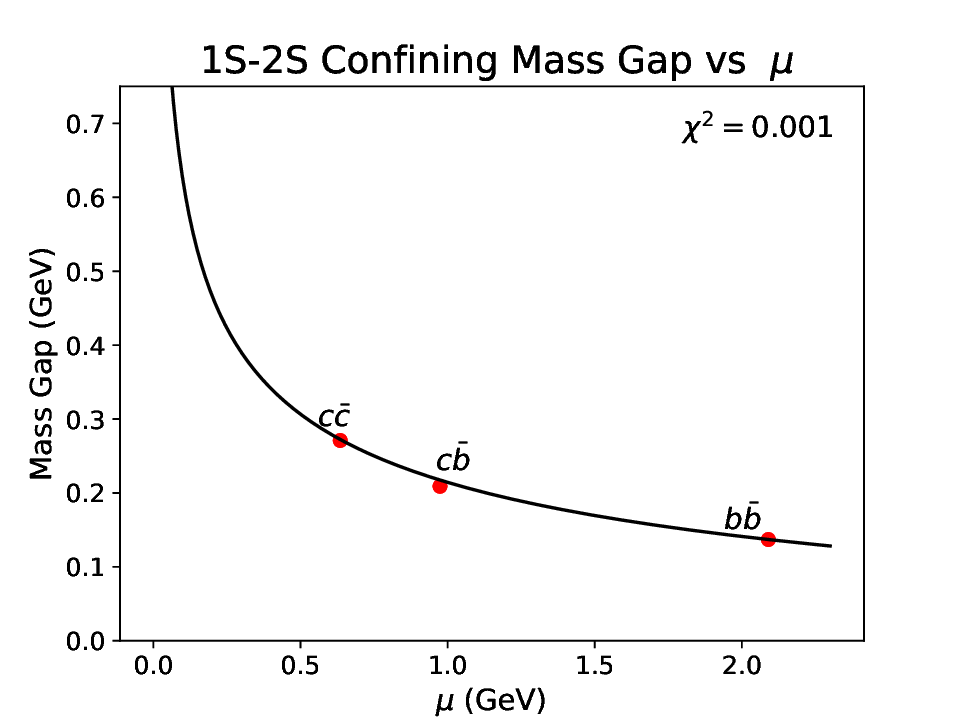}
    \caption{Calculated confining contribution to heavy meson mass gaps vs reduced mass with $a \approx 141$ MeV. Included are mass confinement contributions to heavy-heavy mesons calculated in the light-front \citep{Jafar}. }
    \label{meson graph 2}
\end{figure}

The true model for confinement is still largely in the air. We will apply the quantum theory of the absolute linear potential to model confinement in mesons and explore its implications. Mesons are two-body systems of quark and anti-quark pairs interacting via the confining potential. If we take the confining potential to be linear, we can equivocally think of this system in terms of the reduced mass and treat the system as a single particle in this potential. We restrict our study to heavy mesons, as we are not taking into account relativistic effects, which would be important for lighter systems. Thus, the meson in this work is modeled as a
bound quark and anti-quark system, interacting in an absolute linear potential. The reduced mass of this system is given by
\eq{\mu = \frac{m_1 m_2}{m_1 + m_2},}
where $m_1$ and $m_2$ are the masses of the quark and anti-quark, respectively. Taking Eq. (\ref{Energies}) and setting $m=\mu$, we see
\numeq{E_n = -\left( \frac{\hbar^2 F^2}{2 \mu} \right)^{-1/3}\Airy_z(n + 1) , }
where we set $F$ to 1 GeV/fm \citep{Rohlf}. The confining contribution of 1S-2S meson mass gaps are computed as
\eq{E_1-E_0 = -\left( \frac{\hbar^2 F^2}{2 \mu} \right)^{1/3} (\Airy_z(1)-\Airy'_z(1)).}
We find that confinement contributions depend on the cube root of the reduced mass of the meson. We can predict the confinement contribution to the net mass gap and compare our predictions to those made in the light-front gauge\citep{Jafar}. For comparison with the data, we add a vertical shift parameter $a$ in lieu of the QCD vacuum energy. Doing so results in the following:
\eq{E_1-E_0 = -\left( \frac{\hbar^2 F^2}{2 \mu} \right)^{1/3} (\Airy_z(1)-\Airy'_z(1)) -a.}
Comparing this with calculations done in the light-front quark-model analysis for the confinement effect~\cite{Jafar} gives reasonable predictions in the heavy meson regime with $\chi^{2}=.001$ as seen in Fig. \ref{meson graph 2}, although it tends to deviate as reduced mass decreases. The value of $a$ is calculated by fitting the data to pass through the $b\Bar{b}$ meson point, and upon doing so we find $a \approx 141$ MeV, which interestingly appears consistent with the QCD vacuum energy density range between (163 MeV)$^4$ and (190 MeV)$^4$ discussed recently by Y. Bai and T.-K. Chen~\citep{bai}.

\section{\label{Neutron} Applications to ultra-cold Neutrons \protect}

An interesting application of the one-sided linear model is to the bouncing of ultra-cold neutrons under a gravitational potential. This is done by equating the energy values of the one-sided potential, given by Eq. (\ref{one side energies}), to the potential energy due to gravity near Earth's surface,

\numeq{E_n = -\left(\frac{\hbar^2 F^2}{2 m}\right)^{1/3}\Airy_z(n + 1)=mgh_{n}=Fh_{n} .}
Rearranging this, we find that the heights allowed by the energy eigenstates, $h_n$, are given by 

\numeq{h_{n}= -\left(\frac{\hbar^2}{2 mF}\right)^{1/3}\Airy_z(n + 1) . }
This is in excellent agreement with the predictions made in the WKB approximation.

\begin{table}[h]
\caption{\label{Tab 1} Comparing the analytic energy eigenvalues to the predictions approximated with the WKB method, in micrometers ($\mu m$).}
\begin{ruledtabular}
\begin{tabular}{lccccc}
Source               & $h_1$ & $h_2$ & $h_3$ & $h_4$ & $h_5$ \\
\hline
Exact         & 13.73 & 24.00 & 32.41 & 39.85 & 46.64
\\
WKB Approx. \footnote{These approximations are taken from J. J. Sakurai and J. Napolitano\citep{Sakurai}.}         & 13.63 & 23.98 & 32.40 & 39.85 & 46.65    \\
\end{tabular}
\end{ruledtabular}
\end{table}

The calculated heights can be compared to the data collected by V. V. Nesvizhevsky et al., for which we see the energy eigenvalues modulated by the zeros of the Airy function correspond to jumps in the rate of detection of neutrons at a given height\citep{neutron}. It's also shown that as the height of the neutrons increases, then the rate of detection approaches a classical prediction. This is well explained by the Airy function model as well, as the spacing between Airy function zeros shrinks as $x$ approaches negative infinity, even going to zero in the WKB approximation (shown in Appendix \ref{appen d}). This would correspond to the heights becoming increasingly closer together as the energy eigenstates increase, which means the rate at which neutrons are detected should approach a more classical model.

\section{\label{Conclusion}Conclusion and Outlook\protect}

In this paper, we derived the linear kernel and showed that it was a generalization of the free kernel. We showed the non-dispersive time evolution of the Airy wave packet when released in a free potential by utilizing the Feynman path integral time evolution formulation. We then derived the Airy eigenfunctions of various potentials of linear shape, as well as their corresponding eigenvalues which we can see are modulated by the zeros of the Airy function. These solutions have interesting applications to heavy meson mass spectroscopy and the bouncing ultra-cold neutrons close to Earth's surface. We have shown that the confinement contribution to the heavy meson 1S-2S mass gap is related to the differences in Airy function zeros and agrees well with the confinement effect estimated by the light-front quark-model analysis. This yields the parameter $a$ value associated with the QCD vacuum energy around $a \approx 141$ MeV, which appears consistent with the { recent estimation~\cite{bai} of the QCD vacuum energy density range, (163 MeV)$^4 \sim$ (190 MeV)$^4$}. We have also 
discussed the application to the ultra-cold neutron bouncing measurement which demonstrated a remarkable quantum effect that the allowed heights of ultra-cold bouncing neutrons are quantized by the zeros of the Airy function. 

For future work, we would like to expand our { analysis including the relativistic effect. In particular, it has been noted that the long-distance behavior of the solutions to the linear potential do not go to zero, but instead become oscillatory
in deriving the wave function relativistically
\citep{bag-model}. 
 Such oscillations might be indicative of having enough energy to produce another quark and anti-quark pair in the color flux tube, splitting the meson and into two or more mesons and preserving quark confinement~\citep{confinement}. Thus, further implication of the Airy function solution in  the linear potential may be worthwhile to be explored from the perspectives of the 1+1D QED ``Schwinger model", the 1+1D large $N_c$ QCD ``'tHooft model", the 3+1D light-front quark model and ultimately QCD.}  

\section*{\label{Acknowledgments}Acknowledgments \protect}

We would like to thank 
{ Prof. Robert Golub for useful discussions on the Airy function}. This work was supported in part by the U.S. Department of Energy (Grant No. DE-FG02-03ER41260). 
The National Energy Research Scientific Computing Center (NERSC) supported by the Office of Science of the U.S. Department of Energy 
under Contract No. DE-AC02-05CH11231 is also acknowledged.

\section*{\label{Declarations}Author Declarations \protect}

\subsection*{\label{Conflicts} Conflict of Interest \protect}

The authors have no conflicts to disclose.

\appendix\label{appendix}





\section{Airy Function Derivation \label{appen b}}

The Stokes equation has the form
\begin{equation}
\frac{\partial^2y(x)}{\partial x^2} - xy(x) = 0, \label{appenstokes}
\end{equation}
where we take all variables to be dimensionless. Noting that the definition of the Fourier transform for $y(x)$ is the following
\begin{equation}
    y(x) = \frac{1}{\sqrt{2\pi}} \int_{-\infty}^{\infty} dk  \tilde{y}(k) e^{ikx}.
\end{equation}
Then, we find that 
\begin{equation}
    \frac{\partial^2y(x)}{\partial x^2} = \frac{1}{\sqrt{2\pi}} \int_{-\infty}^{\infty} dk \left[(-k^2) \tilde{y}(k) \right ] e^{ikx},\label{appen1}
\end{equation}
and
\begin{align}
    x y(x) = \frac{1}{\sqrt{2\pi}} \int_{-\infty}^{\infty} dk \left[i \frac{\partial}{\partial k} \tilde{y}(k)\right] e^{ikx}. \label{appen2}
\end{align}
Taking Eq.(\ref{appen1}) and Eq.(\ref{appen2}) and plugging them back into Eq.(\ref{appenstokes}), we find the equation in Fourier space as 
\begin{align}
    \mathcal{F}\left( \frac{\partial^2y(x)}{\partial x^2} - x y(x) \right) &= -k^2 \tilde{y}(k) - i \frac{\partial \tilde{y}(k)}{\partial k} = 0.
\end{align}
Solving the differential equation, we arrive at 
\begin{align}
\tilde{y}(k) = e^{i\frac{k^3}{3}} .   
\end{align}
This leaves us with the Fourier transform Airy $\Airy$ function:
\numeq{y(x) = \mathrm{Ai}(x) = \frac{1}{\sqrt{2\pi}} \int_{-\infty}^{\infty} dk e^{i \frac{k^3}{3}} e^{i k x}.}
In the context of this paper, when working in momentum space we note that $k = px_0 / \hbar$.

\section{Derivation of the Kernel in the Feynman Path Integral Formulation \label{appen e}}

The kernel of a quantum system is defined as 

\begin{align}
    K(x, t; x', t')  & = <x,t|x',t'> \nonumber \\
    & = <x|e^{\frac{-i}{\hbar}H(t-t')}|x'>   .
\end{align}

By slicing Eq. (\ref{kernel def}) into $N$ time steps, it becomes

\begin{align}
    K(x, t; x', t')  & = \bra{x}e^{\frac{-i}{\hbar}H(t-t')}\ket{x'} \nonumber \\ 
    & =\int dx_{N-1} \dots dx_{1} \prod^{N-1}_{i=0}\bra{x_{i+1}}e^{\frac{-i}{\hbar}H\epsilon} \ket{x_{i}}, \label{psum} 
\end{align}
where $ \epsilon=(t-t')/N$.
This calculation will become exact when we take the limit as $N \rightarrow \infty$, as $\epsilon \to 0$ in this limit. Inserting completeness and taking the mean value
$\bar{x} = (x_i + x_{i+1})/2$ between $x_i$ and $x_{i+1}$, we get 
\begin{align}
e^{\frac{-i}{\hbar}V(\bar{x})\epsilon}\int^{\infty}_{-\infty} dp \braket{x_{i+1}|p}\bra{p}e^{\frac{-i}{\hbar}(\frac{p^2}{2m})\epsilon}\ket{x_i}  & = \\ \nonumber e^{\frac{-i}{\hbar}V({\bar x})\epsilon}\frac{1}{2 \pi \hbar}\int^{\infty}_{-\infty} dp e^{\frac{-i}{h}((x_{i+1}-x_{i})p+\frac{\epsilon}{2m}p^2) },
\end{align}
where we used $\braket{p|x_i} = e^{-i p x_i /\hbar}/\sqrt{2\pi\hbar}$ and alike. This integral can then be evaluated by taking advantage of the identity
\numeq{\int_{-\infty}^\infty dx \exp[-i(Ax^2 + Bx)] = \frac{\sqrt{\pi}}{\sqrt{iA}}\exp\left[ \frac{iB^2}{4A} \right],  \label{useful integral}}
and in using the identity, we find that 
\begin{align}
 \bra{x_{i+1}}e^{\frac{-i}{\hbar}H\epsilon}\ket{x_i} = \sqrt{\frac{m}{2\pi i \hbar \epsilon}}e^{\frac{i \epsilon}{\hbar}(\frac{m}{2\epsilon^{2}}(x_{i+1}-x_{i})^2-V(x))}.
\end{align}
Substituting into Eq. (\ref{psum}) gives,

\begin{align}
K(x, t; x', t')  & = \lim_{N \rightarrow \infty}\left (  \frac{m}{2 \pi i \hbar \epsilon } \right )^{N/2}\int dx_{N-1} \dots dx_{1}  \\ \nonumber
& \times \exp \left[\sum_{i=0}^{N-1} \frac{i \epsilon}{\hbar} \left ( \frac{m}{2}\left (\frac{(x_{i+1}-x_{i})}{\epsilon} \right )^2-V(\bar{x}) \right ) \right].
\end{align}

Now, taking the limit as $N$ goes to infinity, we interpret the term in the exponential as a Riemann sum, noting $(x_{i+1}-x_{i})/\epsilon=v(t_i)$ to get the following integral. 

\begin{align}
    K(x, t; x', t')  & =\lim_{N \rightarrow \infty}\left ( \frac{m}{2 \pi i \hbar \epsilon } \right ) ^{N/2}\int dx_{N-1} \dots dx_{1} \\ \nonumber
    & \times \exp\left[ \frac{i}{\hbar}\int_{t'}^{t} dt \left (\frac{m}{2}v(t)^2-V[x(t)] \right )\right],
\end{align}
which gives us the well-known form of the kernel in the Feynman Path Integral,
\begin{align}
K(x,t;x',t')=
\int_{x(t')}^{x(t)} \mathcal{D}[x(t)] e^{\frac{i}{\hbar}S[x(t)] }\label{kernel def},
\end{align}
substituting in 
\eq{\mathcal{D}[x(t)]=\lim_{N \to \infty} (\frac{m}{2 \pi i \hbar \epsilon })^{N/2}dx_{N-1} \dots dx_{1}.}
Since $\hbar$  is small, we can make use of the stationary phase approximation. This states that the smallness of $\hbar$ makes it such that all phases, except the stationary phase, where the action is stable, will oscillate rapidly and destructively cancel. In this case, the stationary phase corresponds to the classical path, so we can rewrite the action term as 
\begin{align}
    S[x(t)] & = S[x_{\text{cl}}(t) + \delta x(t)] \nonumber \\
    & = S[x_{\text{cl}}(t)]+\frac{\partial S}{\partial x}|_{x=x_\text{cl}}\delta x+\frac{1}{2!}\frac{\partial^2 S}{\partial x^2}|_{x=x_\text{cl}}(\delta x)^2+\dots, 
\end{align}
substituting this into our kernel gives
\begin{align}
    K(x,t;x',t') & = e^{\frac{-i}{\hbar} S[x_{\text{cl}}(t)]} \int_{x'}^x \mathcal{D}[x(t)] e^{\frac{i}{\hbar}S[\delta x(t)]}  \nonumber \\
    & = K(\delta x, t; \delta x', t')e^{\frac{i}{\hbar} S[x_{\text{cl}}(t)]} \nonumber \\
    & = K(0, t; 0, t') e^{\frac{i}{\hbar} S[x_{\text{cl}}(t)]}\nonumber \\ 
    & = J(t,t')e^{\frac{i}{\hbar} S[x_{\text{cl}}(t)]} ,\label{appen J}
\end{align}
where the Van Vleck determinant $J(t, t') = K(0, t; 0, t')$ is a purely time-dependent prefactor as $\delta x'= \delta x = 0$ at the endpoints.

\section{Momentum Space Representation of The Linear Kernel \label{appen c}}

We begin with the definition of the kernel in momentum space: 

\begin{align}
    K^{(p)}_{\text{Lin}} & = \bra{p}e^{\frac{-iHt}{\hbar}}\ket{p'} \nonumber \\
    & = \int dx \int dx' \braket{p | x} \bra{x} e^{\frac{-iHt}{\hbar}}\ket{x'}\braket{x' | p'} \nonumber\\
    & = \int dx \int dx' \braket{p | x} K^{(x)}_{\text{Lin}} \braket{x' | p'}, \label{momentum kernel}
\end{align}
where $K^{(x)}_{\text{Lin}}$ has the form of Eq. (\ref{Kx}). We can see that by inserting identity operators, we've recovered the position space kernel inside the double integral. It's also known that 
\numeq{\braket{x | p} = \frac{1}{\sqrt{2\pi \hbar}}e^{-ipx/\hbar} .\label{xp identity} }
In substituting Eq. (\ref{xp identity}) into Eq. (\ref{momentum kernel}), and reintroducing $\Delta t = t-t'$, we see
\begin{align}
    K^{(p)}_{\text{Lin}} & = \frac{1}{2\pi \hbar}  \int dx \int dx' e^{ipx/\hbar} K^{(x)}_{\text{Lin}} e^{-ip'x'/\hbar}  \nonumber\\
    & = \frac{1}{2\pi \hbar}\sqrt{\frac{m}{2\pi i \Delta t\hbar}} \int_{-\infty}^\infty  \int_{-\infty}^\infty   e^{ipx/\hbar} \exp \left[ \frac{im}{2\hbar} \left( \frac{(x-x')^2}{\Delta t} \right. \right.  \nonumber \\ 
    & \left. \left. - \frac{F\Delta t}{m}(x + x') - \frac{F^2}{24m^2}\Delta t^3 \right) \right] e^{-ip'x'\hbar} dx' dx . \label{complicated equation}
\end{align}
After expanding the quadratic $x$ and $x'$ term in the exponential, we perform the integral over $x'$, for which we find
\begin{align}
    & \int_{-\infty}^\infty dx' \exp \left[ \frac{-i}{\hbar} \left( \left( \frac{mx}{\Delta t} + \frac{k}{2}\Delta t + p' \right)x' - \frac{mx'^2}{2\Delta t} \right) \right] \nonumber \\ 
    & = \sqrt{\frac{2i \pi \hbar \Delta t}{m}} \exp\left[ \frac{i}{\hbar} \left( \frac{mx^2}{2\Delta t} + \frac{kx\Delta t}{2} \right. \right. \nonumber\\ 
    & \left. \left. + p'x + \frac{k^2 \Delta t^3}{8m} + \frac{kp'\Delta t^2}{2m} + \frac{p'^2\Delta t}{2m} \right) \right] ,
\end {align}
once again making use of the identity in Eq. (\ref{useful integral}). Upon substituting this back into Eq. (\ref{complicated equation}), we can solve for the linear potential kernel in momentum space:
\begin{align}
    K^{(p)}_{\text{Lin}} & = \exp\left[ \frac{i}{\hbar} \left( \frac{F^2}{12m}\Delta t^3  + \frac{Fp'\Delta t^2}{2m} + \frac{p'^2\Delta t}{2m} \right) \right] \nonumber\\
    &  \times \frac{1}{2\pi \hbar} \int_{-\infty}^\infty dx \exp \left[ \frac{i}{\hbar} \left( -F\Delta t - p' + p \right)x \right] \nonumber \\
    & = \delta\left( p - p' - F\Delta t \right) \exp\left[ \frac{i}{\hbar} \left( \frac{F^2 \Delta t^3}{12m} \right. \right. \nonumber \\
    & \left. \left. + \frac{F p' \Delta t^2}{2m} + \frac{p'^2t}{2m} \right)\right]. \label{momentum prop}
\end{align}
In the limit that $F$ goes to 0, Eq. (\ref{momentum prop}) will become the kernel of the free potential in momentum space, similar to how the linear kernel in position space goes to the free kernel under the same limit.

\section{Airy Function Zero spacing at large $n$ in WKB Approximation \label{appen d}}

As expressed by Sakurai and Napolitano\citep{Sakurai}, the WKB approximated energy eigenstates of neutrons under the influence of gravity is given by 

\numeq{E_{n}= \left\{  \left[ 3(n-\frac{1}{4} ) \pi \right]^{2/3} \right\} \frac{(m g^{2} \hbar^{2})^{1/3}}{2}.}
Taking the difference of the $n$th energy eigenvalue and the preceding $n-1$ eigenvalue gives;

\begin{align}
    &E_{n}-E_{n-1} =\left\{\left[ 3(n-\frac{1}{4} ) \pi \right]^{2/3}  \right. \nonumber \\
    & \left. -\left[ 3(n-\frac{5}{4} ) \pi \right]^{2/3}\right\}\frac{(m g^{2} \hbar^{2})^{1/3}}{2}.
\end{align}
In taking the limit as $n$ approaches infinity, the term in the curly brackets approaches zero, and so the gap between the zeros of the Airy function, and therefore our energy eigenstates $E_{n}-E_{n-1}$ approaches 0.

\nocite{*}
\bibliography{references}

\end{document}